\documentclass[apj,graphicx]{emulateapj}

\begin{document}

\title{Relativistic iron emission lines in neutron star low-mass X-ray binaries \\ as probes of neutron star radii}
\shortauthors{Cackett et al.}
\shorttitle{Fe K lines as probes of neutron star radii}

\author{Edward~M.~Cackett\altaffilmark{1,2}, Jon~M.~Miller\altaffilmark{1}, Sudip Bhattacharyya\altaffilmark{3,4}, Jonathan E. Grindlay\altaffilmark{5}, Jeroen Homan\altaffilmark{6}, Michiel van der Klis\altaffilmark{7}, M.~Coleman~Miller\altaffilmark{4}. Tod E. Strohmayer\altaffilmark{8}, Rudy Wijnands\altaffilmark{7}}
\email{ecackett@umich.edu}
\altaffiltext{1}{Department of Astronomy, University of Michigan, 500 Church St, Ann Arbor, MI 48109-1042, USA}
\altaffiltext{2}{Dean McLaughlin Postdoctoral Fellow}
\altaffiltext{3}{CRESST and Astrophysics Science Division, NASA/GSFC, Greenbelt, MD 20771, USA}
\altaffiltext{4}{Department of Astronomy, University of Maryland, College Park, MD 20742-2421, USA}
\altaffiltext{5}{Harvard-Smithsonian Center for Astrophysics, 60 Garden St, Cambridge, MA 02138, USA}
\altaffiltext{6}{MIT Kavli Institute for Astrophysics and Space Research, 70 Vassar St, Cambridge, MA, 02139, USA}
\altaffiltext{7}{Astronomical Institute `Anton Pannekoek', University of Amsterdam, Kruislaan 403, 1098 SJ, Amsterdam, the Netherlands}
\altaffiltext{8}{Astrophysics Science Division, NASA/GSFC, Greenbelt, MD 20771, USA}

\begin{abstract}

Using {\it Suzaku} observations of three neutron star low-mass X-ray binaries (Ser~X$-$1, 4U~1820$-$30 and GX~349+2) we have found broad, asymmetric, relativistic Fe K emission lines in all three objects.  These Fe K lines can be well fit by a model for lines from a relativistic accretion disk (`diskline'), allowing a measurement of the inner radius of the accretion disk, and hence an upper limit on the neutron star radius.  These upper limits correspond to  $14.5 - 16.5$ km for a 1.4 M$_\odot$ neutron star.  The inner disk radii we measure with Fe K lines are in good agreement with the inner disk radii implied by kHz QPOs observed in both 4U~1820$-$30 and GX~349+2, supporting the inner disk origin for kHz QPOs.  Additionally, the Fe K lines observed in these neutron stars are narrower than those in the black holes that are thought to be close to maximally spinning, as one would expect if inferences for spin are robust.

\end{abstract}


\section{Introduction}
The equation of state of the ultra-dense matter in neutron stars remains enigmatic, however, it is through accurate measurements of neutron star radii and masses that it can be constrained \citep[e.g.][]{latt_prak07,klahn06}.  Unfortunately, there are very few accessible and robust observational measurements of neutron star radii.

There are currently several methods that utilize X-ray spectroscopy/timing to determine neutron star radii.  For instance, if one takes the origin of the kHz quasi-period oscillations (QPOs) to be associated with the inner edge of the accretion disk then this places constraints on mass and radius \citep{miller98}.  However, while these kHz QPOs can be measured precisely, and are likely associated with the inner accretion disk, their origin is not certain \citep{lamb01,stella99,abramowicz03,titarchuk02}.  Burst oscillations during thermonuclear X-ray bursts may be used to constrain the neutron star radius \citep{bhattacharyya05}, but this method alone cannot provide very tight constraints as there are many unknown parameters.  Alternatively, thermal continuum emission from the stellar surface can give a measure of radius \citep[e.g.][]{vanparadijs87,heinke06,zavlin07,ho07,bogdanov07} but suffer from distance and/or model uncertainties.  Additionally, X-ray absorption lines from the stellar surface would measure the gravitational redshift, but these lines appear to be extremely rare and difficult to observe, and the only detection is of modest statistical significance \citep{cottam02,kong07}. In this paper we present a method for placing upper limits on neutron star radii using relativistic iron emission lines \citep[see][for a review]{miller07} that originate in the inner accretion disks around the neutron stars in X-ray binaries.  

Velocity shifts encoded into disk emission lines directly reflect the radius at which the line is formed.  In the X-ray spectra of accreting black holes and neutron stars, iron K-shell emission lines are widely observed \citep{nandra97,reynolds97,white85,white86,hirano87,asai00}.  The extreme red-wing of some iron emission lines implies that they are formed in the inner accretion disks and are primarily shaped by dynamics including relativistic Doppler shifts, due to the high velocities in the disk, and gravitational redshifts (hence the lines are referred to as being `relativistic'). Therefore, iron K lines can serve as incisive measures of the inner disk radius \citep{fabian89}.

These relativistic, asymmetric Fe K emission lines are well-known and well-studied in the X-ray spectra of both supermassive black holes in Active Galactic Nuclei \citep{nandra97,reynolds97} and stellar-mass black holes in X-ray binaries \citep{miller04}.  In the case of neutron stars, iron emission lines are weaker, however, and until lately prior observations have not clearly revealed a relativistic line profile, and the lines could be well fit by just a Gaussian \citep[e.g.,][]{white86,asai00,oosterbroek01,disalvo05}.  Only very recently have observations of Serpens X-1 with {\it XMM-Newton} shown a relativistic, asymmetric line in a neutron star low-mass X-ray binary (LMXB) for the first time \citep{bhattacharyya07}, confirming the inner disk nature of the lines in neutron star LMXBs.  We note that though the iron line in Serpens X-1 has been observed before \citep{white86,asai00,oosterbroek01}, the new observations by \citet{bhattacharyya07} were the first time a clear asymmetry has been seen in the line profile.

The high effective area of the {\it Suzaku} X-ray telescope \citep{mitsuda07} in the Fe K band combined with its broadband energy coverage, means that it is optimized for studies of inner disks and relativistic regimes using iron lines.  Here we present {\it Suzaku} observations of the Fe K lines in three neutron star LMXBs (\object{Serpens X-1}, \object{4U 1820-30} and \object{GX 349+2}).  We find them all to show a relativistic asymmetric profile.  In section \ref{sec:analysis} we describe the data analysis and results, and in section \ref{sec:disc} we discuss the use of these lines as probes of neutron star radii, as well as comparing the results with kHz QPOs seen in these sources and with iron lines in black holes.  Finally, we summarize our findings in section \ref{sec:conc}.

\section{Data Analysis and Results}\label{sec:analysis}

The data were obtained with the {\it Suzaku} X-ray telescope, which has several detectors on-board.  The X-ray Imaging Spectrometer (XIS) detectors provide spectra from 0.5 - 10 keV, whilst the Hard X-ray Detector (HXD) PIN camera provides spectra at higher energies from 12-40 keV (with current calibrations).  For observations of Serpens X-1 and 4U 1820-30, {\it Suzaku} was operated at the XIS nominal pointing position, with the XIS detectors operating in 1/4 window mode and 1-sec burst clock mode to prevent pile-up.  The observations of GX 349+2 were obtained from the {\it Suzaku} data archive.  Two observations were performed of approximately 25 ksec each, with the telescope pointing at the HXD nominal position.  The XIS was operated with 1/8 window mode and 0.3-sec burst clock mode.  As the observations are only 4 days apart and the source count-rate and spectral shape did not vary significantly, we add the spectra from both observations together to improve the signal-to-noise.  The observations are summarized in Table \ref{tab:summary}.

\begin{deluxetable*}{llcc}
\tablecolumns{4}
\tablewidth{0pc}
\tablecaption{Summary of {\it Suzaku} observations}
\tablehead{
Object & Observation date & \multicolumn{2}{c}{Exposure time of spectrum (ksec)} \\
 & & XIS & PIN}
 \startdata
 Serpens X$-$1 & 24 October 2006 & 30.5 & 27.5\\
 4U 1820$-$30 & 14 September 2006 & 17.9  &  28.4\\
 GX 349+2   &  14 March 2006  &   25.3  & 17.3\\
  & 19 March 2006   &    28.1  &    22.9
\enddata
\label{tab:summary}
\end{deluxetable*}

The standard data spectral products for the XIS were used for all observations.  We also used the canned response matrices, as custom responses cannot currently be created with the burst mode.  For the HXD/PIN data, we extracted the spectrum from the cleaned PIN event files following the standard analysis threads.  The PIN non-X-ray background was extracted  from the time-dependent, observation specific model provided by the instrument team, and was combined with the standard model for the cosmic X-ray background.

We fit the spectra using XSPEC v.11 \citep{arnaud96}.  The continuum fit is performed from $1 - 9$ keV in the XIS and $12-25$ keV in the HXD/PIN. We ignore below 1 keV as this is the region where the accumulation of some carbon-rich material onto the CCDs has changed the effective are curve, additionally we ignore the $1.5-2.5$ keV region where there are large calibration uncertainties due to an instrumental feature and the $4 - 7.5$ keV region where the broadened iron line profile is present.  We only fit the PIN spectrum up to 25 keV as this is the energy at which the background dominates over the source flux for these objects.  While the XIS consists of four separate detectors, we only fit XIS 2 and XIS 3 as their calibrations appear to be most robust.  The XIS 2, XIS 3 and PIN are all fit simultaneously.  The broadband energy coverage provided by the combination of both the XIS and PIN allows an accurate determination of the continuum shape either side of the Fe K band.  This is vital in robustly determining the line profile.

It is well known that neutron star spectra can be fit by a wide variety of different models \citep[e.g.][]{lin07}.  For simplicity, we fit a model with two thermal components (multicolor disk blackbody + blackbody) as per a recent study \citep{lin07}.  {In that study as the sources vary in luminosity, the measured temperatures of both thermal components in this model are found to follow $L \propto T^4$, suggesting a strong physical motivation, where as in other models the measured temperatures of the thermal component do not follow $L \propto T^4$ as the source luminosity increases.  An additional power-law component is also required \citep[as is also found in][]{lin07}, and we note here that all 3 spectral components are required. The spectral index of the power-law is allowed to vary between the XIS 2 and XIS 3 as this is required to remove some slight differences in slope between the calibration of the different detectors.  The power-law component for the PIN is tied to the XIS 2 value, and this is the value quoted in the Table \ref{tab:contfits}.  The parameters from the continuum spectral fits for each object are given in Table \ref{tab:contfits} and the spectral fits to Ser~X$-$1 are shown in Fig. \ref{fig:spectra} (spectral fits to the other objects are similar).  All uncertainties quoted throughout the paper are at the 90\% confidence level.

\begin{figure}
\includegraphics[angle=270,width=8.4cm]{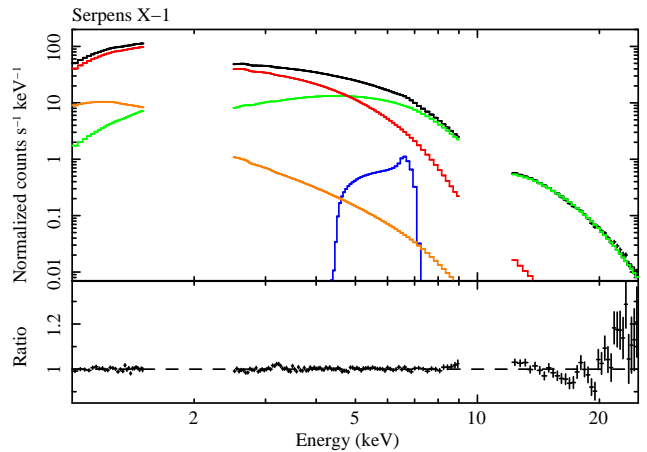}
\caption{Suzaku spectrum of Serpens X-1.  For clarity only the XIS2 (1 - 9 keV) and PIN (12 - 25 keV) data are plotted, and the XIS2 data have been re-binned.  The black line shows the overall model, the colors distinguish the separate model components.  Green is the blackbody, red the multicolor disk blackbody, orange the power-law and blue the diskline.}
\label{fig:spectra}
\end{figure}

\begin{deluxetable*}{llll}
\tablecolumns{4}
\tablewidth{0pc}
\tablecaption{Parameters of the continuum models}
\tablehead{
Parameter & Ser X$-$1 & 4U 1820$-$30 & GX 349+2}
\startdata
$N_H$ (10$^{22}$ cm$^{-2}$  & $0.56 \pm 0.01$  & $0.26\pm0.01$ & $0.93\pm0.01$\\
Blackbody (keV)  & $2.28 \pm 0.02$  & $2.46\pm0.03$ & $2.22\pm0.01$\\
B-body normalization (10$^{-2}$)  & $4.9 \pm 0.2$ &  $3.2\pm0.1$ & $9.7\pm0.1$\\
Disk blackbody (keV) & $1.21 \pm 0.01$ &  $1.15 \pm 0.02$ & $1.29\pm0.01$ \\
Disk b-body normalization   & $103 \pm 5$  &  $55\pm 3$ & $130\pm5$\\
Power-law index  & $3.6\pm0.4$  &  $2.1\pm0.1$ & $1.4\pm0.1$ \\
Power-law norm. (10$^{-2}$) & $18\pm3$ &  $8.17\pm0.02$ & $2.52\pm0.03$\\
Flux\tablenotemark{a} ($10^{-9}$ erg cm$^{-2}$ s$^{-1}$) & $5.9\pm0.3$ & $3.6\pm0.2$ & $10.2\pm0.5$ \\
$\chi^2_\nu$ (d.o.f.) & 1.09 (1936) & 0.99 (1936) & 1.17 (1936)
\enddata
\tablenotetext{a}{Flux is evaluated in the 0.5-10 keV range}
\label{tab:contfits}
\end{deluxetable*}

Fitting the continuum reveals significant iron lines in all three objects, which are all detected at greater than the 7-$\sigma$ level in all objects (which we determine from an f-test).  We show the lines in Fig. \ref{fig:felines}.  Importantly, the lines in these objects are all clearly revealed as having broad, asymmetric profiles.  While we tried fitting a variety of different continuum models, the model described above is the only model that successfully fits both the XIS and PIN data.  However, various other models, such as a disk blackbody + power-law or a blackbody + power-law model give acceptable fits to the XIS data alone.  It is important to note the iron line profile recovered from these different continuum models is robust and changes very little in shape.  This demonstrates that the line is not due to continuum components crossing in the iron band.

\begin{figure}
\centering
\includegraphics[width=8.2cm]{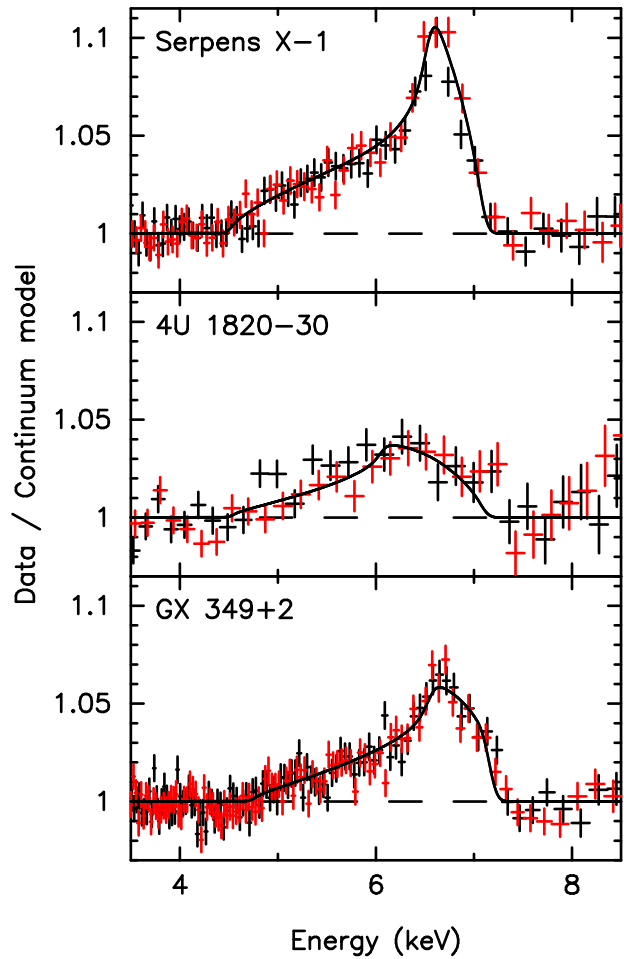}
\caption{Iron lines in the three neutron star LMXBs observed with {\it Suzaku}.  The ratio of count rate to continuum model is shown versus energy.  Black points are for data from the XIS 2 detector, and red points are from the XIS 3 detector.  The solid line is the best-fitting diskline model.}
\label{fig:felines}
\end{figure}

For slowly rotating neutron stars, the gravitational potential around a neutron star is expected to be very close to a simple Schwarzschild potential.  In fact, the neutron star spin for 4U~1820$-$30 and GX~349+2 inferred from the difference between the freqencies of the kHz QPOs is less than 300 Hz, which would imply that the spin parameter $j\equiv cJ/GM^2 < 0.3$ \citep{miller98}.  Frame-dragging corrections for such spins should give errors of much less than 10\% \citep{miller98}.  We therefore fit the iron line profiles with a model of line emission from a relativistic accretion disk assuming the Schwarzschild metric \citep[the `diskline' model,][]{fabian89}.  The line profiles in all objects can be well fit by this model.  When fitting the lines we restrict the line energy to between $6.4-6.97$ keV (the allowed range for different ionization states of Fe K emission).  The inner radius, inclination and disk emissivity are free parameters in the fit, whilst the outer disk radius is fixed at a large value (1000 $R_G$).

The inner radius, $R_{in}$, of the accretion disk (in units of gravitational radii, ${\mathrm R_G  = GM/c^2}$) is important for shaping the line, and is a parameter of the line fit. We are therefore able to measure the inner radius of the accretion disk in the three objects.    The full `diskline' parameters are given in Table \ref{tab:linefits}.  The inner radius of the accretion disk places an upper limit on the neutron star radius.  In general, we find that these inner disk radii are in the range $7 - 8~{\mathrm R_G}$, which is equivalent to $14.5 - 16.5$ km for a 1.4 M$_\odot$ neutron star.

\begin{deluxetable*}{llll}
\centering
\tablecolumns{4}
\tablewidth{0pc}
\tablecaption{Diskline parameters and kHz QPO details}
\tablehead{Parameter & Serpens X$-$1  & 4U 1820$-$30 & GX 349+2}
\startdata
Line energy (keV) & $6.83^{+0.15}_{-0.06}$ &  $6.97_{-0.18}$ & $6.97_{-0.02}$\\
$R_{in} (GM/c^2)$  & $7.7\pm0.5$   &  $6.7^{+1.4}_{-0.7}$  & $8.0\pm0.4$\\
Emissivity index, $\beta$   & $-4.8^{+0.4}_{-0.8}$  &  $-4.3^{+1.1}_{-2.1}$ & $-4.1\pm0.3$\\
Inclination (degrees)  & $26 \pm 2$ &  $19^{+5}_{-18} $  & $23\pm1$\\
Normalization ($10^{-3}$) & $6.6 \pm 0.6$  &  $1.8\pm0.4$   & $7.6\pm0.6$ \\
Equivalent width (eV) & $ 132 \pm 12$ &  $51\pm11$   & $76\pm6$\\
$\chi^2_\nu$ (d.o.f.) & 1.05 (3851) & 1.00 (3851) & 1.06 (3851)\\
\hline
$R_{in}$ (km) from Fe K line & $15.9\pm1.0$  & $13.8^{+2.9}_{-1.4}$ &  $16.5\pm0.8$\\
Upper kHz QPO (Hz)  & -  & $1060\pm20$  &  $978\pm9$\\
$R_{in}$ (km) from kHz QPO  & - & $16.1\pm0.2$ &  $17.0\pm0.1$
\enddata
\tablecomments{The $R_{in}$ measurements given in km are evaluated at M = 1.4 M$_\odot$. Ser X$-$1 does not have a detected kHz QPO.  kHz QPO measurements are taken from \citet{zhang98a} for 4U~1820$-$30, where we choose the value where the upper kHz QPO frequency reaches a maximum and remains constant with increasing count rate.  For GX~349+2 we use the kHz QPO measurement from \citet{zhang98b}.}
\label{tab:linefits}
\end{deluxetable*}

We checked that the `diskline' is the appropriate line model to use.  Alternative models have been developed for emission lines formed in the space-time around a spinning black hole.  We find that fits with the `laor' model \citep{laor91} for a maximally spinning black hole give inner disk radii that are consistent with the diskline fits.  We chose the diskline model since reasonable expectations for neutron star parameters suggest a spin parameter $< 0.3$.

In addition to fitting with the diskline model, it is also possible to put a lower limit on the gravitational redshift, $z = GM/Rc^2$, from fitting the asymmetric line profiles with two Gaussians, one narrow to fit the peak, and one broad to fit the red wing.  We compare the measured center energy of the broad Gaussian component with the line rest energy.  If the disk extends all the way down to the neutron star surface, one could naively interpret the energy shift as the surface redshift.  Doing this we get redshifts in the range 0.1$-$0.2, providing a consistency check with `diskline' fits.

We also note that there is expected to be a `boundary layer' between the inner accretion disk and the neutron star. This material is expected to be optically thick, and not in a Keplerian orbit around the star.  The blackbody component of our continuum model can be interpreted as this component.  In alternative continuum models this is often modelled as a Comptonized component, though the high effective optical depth means that it essentially behaves as a blackbody.
In neutron star spectra there does appear to be two thermal components that follow $T^4$ as the source luminosity increases \citep{lin07}, one a Keplerian accretion disk and the other a small, optically-thick boundary layer.  Moreover, the line profiles we observe are well fit by the `diskline' model, suggesting that such a boundary layer does not strongly affect the line profile.  In fact, the inner disk radii that we measure may imply that the boundary layer is small, for reasonable neutron star radii.  Additionally, the frequencies of kHz QPOs observed in neutron star systems are likely associated with the orbital frequency of the inner disk, and the frequency of this QPO is similar to the expected Keplerian frequency there \citep{miller98}, indicating that the inner disk is likely in Keplerian orbits.

Finally, we note that to achieve asymmetric lines are observed here the iron line emission must originate in the accretion disk.  Alternative models for line broadening, such as a Comptonizing coronae would require a very high optical depth, making the Comptonizing material effectively a disk \citep{reynolds00}. 

\section{Discussion} \label{sec:disc}

{\it Suzaku} observations of Ser~X$-$1, 4U~1820$-$30 and GX~349+2 have revealed broad, relativistic Fe K emission lines in each of the objects.  By fitting with the `diskline' model, we were able to measure the inner radius of the accretion disk, and hence an upper limit on the neutron star radius.  Only in one case previously has a clearly asymmetric line profile been observed \citep{bhattacharyya07}, and that was in Ser~X$-$1 with {\it XMM-Newton}.  The iron line that we see in Ser~X$-$1 with {\it Suzaku} is very similar in shape to that seen independently with {\it XMM-Newton} by \citet{bhattacharyya07}, and the inner disk radii are similar, though not quite consistent within 1-$\sigma$.  The observations presented here are more sensitive and have a broader continuum coverage, allowing tighter constraints on the inner disk radius. The fact that relativistic content is now clear in the independent work of \citet{bhattacharyya07} and in the more sensitive {\it Suzaku} spectra presented here, signals that these lines are robust and can now be used to constrain the nature of neutron stars.

In Fig. \ref{fig:eos} we show the constraints on radius and mass from these new {\it Suzaku} observations.  They are completely independent of any previous constraints and rely on different physics.  For comparison we include the constraints from kHz QPOs observed in X-ray binaries. We also show mass-radius curves from a variety of equations of state.  The curves shown are not meant to be exhaustive, but are just shown to demonstrate a range in mass-radius expected from different equations of state.  Whilst the radius upper limits from our current observations cannot rule out any equations of state, the potential of using these lines can be seen.  Moreover, if masses for these neutron stars are determined, then the region of allowed mass-radius space is significantly reduced. 

\begin{figure}
\centering
\includegraphics[width=8.2cm]{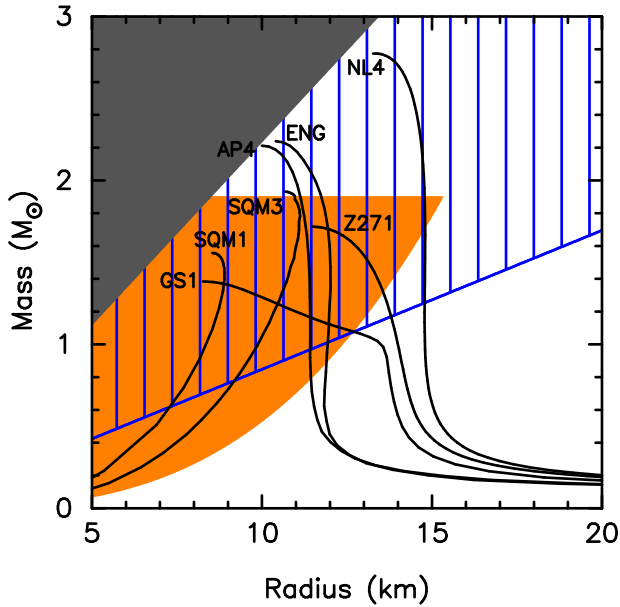}
\caption{Neutron star gravitational mass and radius constraints.  The gray region is ruled our because of causality (the sound speed must be less than the speed of light).  The orange region is the allowed region from the highest frequency QPO observed \citep[1330 kHz in 4U~0614+091,][]{vanstraaten00} and assuming the neutron star spin, $cJ/GM^2 = 0.2$, where $J$ is the stellar angular momentum.  The blue region is allowed from this work on neutron star iron lines - here we plot the most well-constrained upper limit, which is from GX~349+2.  For comparison, we also plot the mass-radius curves for a variety of equations of state.  All curves are as labelled in \citet{latt_prak01} (see references therein for details of the equations of state), except for NL4 which is from \citet{akmal98} and Z271 which is from  \citet{horowitz01}.}
\label{fig:eos}
\end{figure}

Indeed, our analysis of broad iron lines in two sources that exhibit kHz QPOs \citep[4U~1820$-$30, GX~349+2;][]{zhang98a,zhang98b} opens up a new way to estimate the masses of the neutron stars in these systems.  The upper kHz QPO frequency is thought to be close to the orbital frequency at the inner edge of the optically thick emission \citep[e.g.][]{miller98}, and thus from a radius close to the $R_{in}$ inferred from the iron line.  In Schwarzschild coordinates the expressions for orbital speed and frequency measured at infinity are the same as their Newtonian counterparts, $v_{\rm orb}=\sqrt{GM/r}$ and
$\nu_{\rm orb}=(1/2\pi)\sqrt{GM/r^3}$.  Therefore, a simultaneous measurement of speed and frequency allows a measurement of the stellar mass regardless of the radius:
\begin{equation}
M={v_{\rm orb}^3\over{2\pi G\nu_{\rm orb}}}\; .
\end{equation}
Frame-dragging corrections to this expression for a spacetime with angular momentum parameter $j\equiv cJ/GM^2$ are only of order $\sim j/(r/M)^{3/2}$ \citep[e.g.,][]{markovic00}, which is less than 2\% for the expected $j<0.3$.  Simultaneous spectral and timing observations of 4U~1820--30 would be especially valuable due to the tentative inference of a mass $M>2\,M_\odot$ from a saturation of QPO frequency vs. count-rate and color (Zhang et al. 1998a; Bloser et al. 2000; see M\'endez 2006 for a dissenting view).

Pending such a coordinated analysis, in Table~3 we present a simplified consistency test between the timing and spectral data in 4U~1820$-$30 and GX~349+2.  For a given QPO frequency, $\nu$, the radius is bounded from above by $R_{\rm max}=\left[GM/(4\pi^2\nu^2)\right]^{1/2}$, hence for a given source the largest observed frequency yields the smallest upper limit $R_{\rm max}$.  We list these limits in Table~\ref{tab:linefits}, assuming a mass $M=1.4\,M_\odot$.  The general consistency between spectral and timing limits is encouraging.  We also see that the best values for 4U~1820$-$30 are somewhat offset, and find that the agreement would be improved for higher mass (e.g., $M=2\,M_\odot$ works), but the current broad errors on the iron line and lack of simultaneous timing measurements prevent us from drawing conclusions at this time.

We note that we find low values ($<30^\circ$) for the system inclinations in all cases.  This may be a selection effect as the observed objects were chosen based on strong lines seen in previous {\it ASCA} data \citep{asai00}, and the strongest lines are present in systems with the lowest inclinations.  Ser~X$-$1 and GX~349+2 do not have previously well defined inclinations, though Ser~X$-$1 is expected to be $<60^\circ$ as no energy dependent X-ray dips or eclipses are seen \citep{frank87} and GX~349+2 is also suggested to have a low inclination \citep{wachter96,oneill01}.  For 4U~1820$-$30, an inclination of 35 - 50$^\circ$ was determined from UV observations \citep{anderson97}, and observations of the X-ray superburst find limits that are consistent with this \citep{ballantyne04}, though also allow slightly lower inclinations.  Differences in the inclination may be reconciled if the inner disk is at a different inclination to the binary system.  This may be possible as radio jets are not  always seen to align with the binary inclination.  For example, the radio jet in the black hole X-ray binary GRO~J1655-40 \citep{hjellming95} is inclined at a different angle that the inclination binary system determined from optical lightcurves \citep{orosz97,greene01}.

The presence of relativistic iron lines in neutron stars as well as in black holes suggests that despite the physical and observational differences between the two types of objects, the accretion disks must be fairly similar.  Extremely broad iron lines in some black hole sources are thought to be due to spinning black holes allowing the inner disk to extend further in \citep[e.g.][]{miller04,brenneman06}.  The iron lines we observe in these neutron stars strengthen this argument as the neutron star iron lines all have inner disk radii much greater than these extreme cases in black holes.  

Finally, in observations of iron lines in black holes, a strong reflection component, or `hump', is generally observed above 10 keV \citep[e.g.][]{miniutti07}.  In our observations we do not find strong evidence for such a reflection component.  Currently, reflection spectra  have only been calculated for an ionizing spectrum that is power-law in nature.  We find, though, that neutron star spectra turn over and drop off faster than a power-law and so any reflection hump would be less apparent. 

\section{Conclusions} \label{sec:conc}

We have found broad, relativistic Fe K emission lines in 3 neutron star LMXBs (Ser~X$-$1, 4U~1820$-$30 and GX~349+2) using {\it Suzaku}.  From `diskline' fits to the lines, we were able to measure the inner radius of the accretion disk, and hence place an upper limit on the neutron star radius.  However, present measurements cannot currently rule out any reasonable equations of state.

We found that the inner disk radii we measure with iron lines are consistent with the inner disk radii implied by kHz QPOs in 4U~1820$-$30 and GX~349+2, supporting the inner disk nature of kHz QPOs.  Additionally the iron lines observed in these neutron stars are narrower than those in the black holes that are thought to be spinning, as one would expect if the spin interpretation is correct.

Recently, new observations are yielding important progress in understanding neutron stars, such as constraining the allowed properties of the crust \citep[e.g.][]{cackett06,watts06,stroh_brown02}.  But, determining the equation of state of matter in the core still remains a difficult problem.  As currently one single method will not conclusively determine the equation of state, it is critical to have multiple, independent methods to measure masses and radii of neutron stars.  The advantage of using Fe K emission lines as a probe of the neutron star radius is that they only require short observations to clearly reveal the relativistic lines, and does not require any knowledge of the distance to the object.  Additionally, there are likely to be many neutron star low-mass X-ray binaries that have these lines \citep[e.g.][]{asai00}.  Thus, using relativistic iron lines in neutron stars offers a promising, complementary method to constrain neutron star radii.  Future proposed X-ray observatories such as {\it Constellation-X} and {\it XEUS} will be able to place extremely tight limits on even weak lines.

\acknowledgements
We thank A. Steiner and J. Lattimer for very kindly providing the mass-radius curves for various equations of state.  JMM acknowledges funding from NASA through the {\it Suzaku} GO program.

\end{document}